\documentclass[aps,twocolumn,superscriptaddress,showpacs,floatfix]{revtex4}
\usepackage[bf]{caption}
\usepackage{graphicx}
\usepackage{amsmath}
\usepackage{wrapfig}

\begin{document}

\title{Dynamics and thermodynamics of fragment emission from excited sources}
\author{M.J. Ison}
\affiliation{ Departamento de F\'{i}sica, Facultad de Ciencias Exactas y Naturales, Universidad de 
Buenos Aires,
 Pabell\'on $I$, Ciudad Universitaria, Nu\~{n}ez, $1428$,\\
 Buenos Aires, Argentina.}
\author{C.O. Dorso}
\affiliation{ Departamento de F\'{i}sica, Facultad de Ciencias Exactas y Naturales, Universidad 
de Buenos Aires,
 Pabell\'on $I$, Ciudad Universitaria, Nu\~{n}ez, $1428$,\\
 Buenos Aires, Argentina.}

\date{\today}

\begin{abstract}

In this paper we study the process of fragmentation of highly excited Lennard-Jones drops by 
introducing the concept of emitted fragments (clusters recognizable in configuration space which
live more than a minimum lifetime). We focus on the dynamics and thermodynamics of the emitted 
sources, and show, among other things,  that this kind of process can not be cast into a sequential 
or simultaneous one, and how a local equilibrium scenario comes up, allowing us to define and 
explore a local temperature, which turns out to be a strongly time-dependent quantity.   

\end{abstract}

\pacs{25.70.Mn, 25.70 -z, 25.70.Pq, 02.70.Ns}

\maketitle

\section{Introduction}

The problem of fragmentation attracts the interest of physicists in many branches of physics. 
In particular, in this problem we have to deal with systems that are both finite and 
non-extensive. Moreover, we can face from extremely small systems like the nucleus to huge 
ones like galaxies. In both cases the size of the system under analysis is of the order of the 
range of the interaction potential and calls for a careful analysis. In what follows we will 
focus on the analysis of small Lennard-Jones ($LJ$) drops i.e. a system in which the dynamics 
is generated by a Hamiltonian in which there is a very short-range strong repulsion and a 
short-range attraction. As such we consider that it will give relevant insight into the 
problem of fragmentation.  

This problem has been analyzed from different points of view and we classify them in two main 
groups i.e. Statistical models and dynamical models. In the first case \cite{randrupkooning} 
\cite{smm}\cite{fai} the main assumption is that the fragmentation process is driven by 
phase-space occupancies. The always present assumption of {\it freeze-out} volume states that
the system fragments in an equilibrium scenario at a given fixed volume and that thermal and 
chemical equilibrium is reached, in correspondence with the flavour chosen by the researcher 
(i.e. microcanonical, canonical or grand-canonical). Recently this kind of analysis has been 
extended to the isobaric isothermal ensemble \cite{gulmiflow}. Even in this kind of formulation 
there have been some efforts to include non-equilibrium effects, like for example the presence of
 expansive collective motion \cite{subal}. 

The other kind of models we dubed, dynamical ones, include among others: Classical ones 
\cite{ale97,ale99,noneqfrag}, quantual ones \cite{amd,qmd}, which are fully microscopic. 
In this category we 
also include those based on numerical realization of kinetic equations like BUU, VLD, etc. 
Focusing on the fully microscopic models the main advantage we find is that we have correlations of 
all orders at all times. Moreover non-equilibrium features of the process can be readily explored.

Continuing with a series of previous works we will study the dynamics of highly excited 
$147$ particles $LJ$ drops \cite{ale99}. Previously we have mainly considered the problem 
of fragment formation. We have shown that this phenomena takes place in phase space and as such 
it is not an experimental observable. 
In this communication we will focus our attention on the properties of the emitting sources and
the corresponding emitted fragments. At this point it is worthwhile to define what we understand
as emission.
We say that a fragment has been emitted when it is recognizable in configuration 
space. We have already made this classification \cite{ale97b}, and we have shown that there 
are well defined time scales for the processes of fragment formation and fragment emission.

In Section $II$ we review the different fragment recognition algorithms currently in use. In 
Section $III$ we present different Temperature definitions that will be relevant in the analysis
of the evolution of the excited drops. Section $IV$ deals with the model we study. Section $V$ 
includes all the results we have obtained in our numerical simulations (fragment mass 
distributions, characteristic times, temperature of the emitting sources, role of the radial 
flux, etc). Finally conclusions will be drawn.

\section{Fragment recognition algorithms}

The problem of analyzing molecular dynamics calculations is an old one and is not completely 
settled. To our knowledge there are three main fragment recognition algorithms in use: 
$MST$, $MSTE$, and $ECRA$.

The simplest and more intuitive cluster definition is based on
correlations in configuration space: a particle $i$ belongs to a
cluster $C$ if there is another particle $j$ that belongs to $C$
and $|{\bf r_i}-{\bf r_j}| \leq r_{cl}$, where $r_{cl}$ is a
parameter called the clusterization radius. If the interaction
potential has a cut off radius $r_{cut},$ then $r_{cl}$ must be
equal or smaller than $r_{cut}$, in this work we chose
$r_{cl}=r_{cut}=3\sigma$. The algorithm that recognizes these
clusters is known as the ``Minimum Spanning Tree'' (MST). The main
drawbacks of this method is that only correlations in {\bf
q}-space are used, neglecting completely the effect of momentum.

An extension of the MST is the ``Minimum Spanning Tree in Energy
space'' (MSTE) algorithm \cite{campimste}. In this case, a given a set of particles
$i, j,..., k$, belongs to the same cluster $C_i$ if:
\begin{equation}\forall \, i \, \epsilon \, C_i \:,\: \exists \, j \, 
\epsilon \, C_i  \, /
\,  e_{ij} \leq 0
\end{equation}
where $e_{ij} = V(r_{ij}) + ({\bf p}_i - {\bf p}_j)^2 / 2 \mu$,
and $\mu$ is the reduced mass of the pair $\{i,j\}$.  MSTE
searches for configurational correlations between particles
considering the relative momenta of particle pairs. In spite of
not being supported by a physically-sound definition of a cluster,
the MSTE algorithm typically recognizes fragments earlier than
MST. Furthermore, due to its sensitivity in recognizing promptly
emitted particles, it can be useful to study the
pre-equilibrium energy distribution of the participant particles.

A more robust algorithm is based on the ``Most Bound Partition''
(MBP) of the system~\cite{ecra}.  The MBP is the set of
clusters $ \{ C_i \}$ for which the sum of the fragment internal
energies attains its minimum value:
\begin{eqnarray}
{ \{C_i\} \atop {}}   & { = \atop {}} &  { \hbox{argmin} \atop {\scriptstyle 
\{C_i\}}  } { \textstyle {[E_{ \{C_i\}} = \sum_i E_{int}^{C_i}]} \atop {} } 
\nonumber \\
E_{int}^{C_i}& = & \sum_i[\sum_{j \in C_i} K_j^{cm} + \sum_{ {j,k
\in C_i} \atop j \le k} V_{j,k}] \label{eq:eECRA}
\end{eqnarray}
where the first sum in (\ref{eq:eECRA}) is over the clusters of
the partition, $K_j^{cm}$ is the kinetic energy of particle $j$
measured in the center of mass frame of the cluster which contains
particle $j$, and $V_{ij}$ stands for the inter-particle
potential. It can be shown that clusters belonging to the MBP are
related to the most-bound density fluctuation in {\bf r-p}
space~\cite{ecra}.

The algorithm that finds the MBP is known as the ``Early Cluster
Recognition Algorithm'' (ECRA).  Since ECRA searches for the
most-bound density fluctuations in {\bf q-p} space, valuable space
and velocity correlations can be extracted at all times, specially
at the very early stages of the evolution. This has been used
extensively in many fragmentation
studies~\cite{ale97,ale98,ale99,ecra}
and has helped to discover that excited drops break very early in
the evolution.

When we use the three above-mentioned algorithms and we apply a criterium based on the 
average microscopic stability of the clusters (see for example \cite{ale97}) to determine 
the corresponding times of fragment formation three time scales emerge, which satisfy the 
following relation: $\tau_{ECRA}<\tau_{MSTE}<\tau_{MST}$.
But the meaning of each of these times is quite different. $\tau_{ECRA}$ refers to that time 
at which, on average, clusters attain microscopic stability regardless of the structure of 
the system in $q-space$. $\tau_{MSTE}$ 
has a rather obscure meaning because $MSTE$ algorithm is not well defined from the physical 
point of view for dynamical problems. Finally $\tau_{MST}$ refers to 
that time at which, on average, microscopic stability is attained for free fragments. Please 
take into account that depending on the definition of $r_{cl}$ this last condition can be 
rephrased as "weakly interacting fragments".  

\section{Temperatures}
\label{sectemp}

Because we are aiming to perform some kind of thermodynamic analysis of the system, a 
prescription for the calculation of the temperature in small (and probably out of 
equilibrium) systems, is to be given. 
We should keep in mind that we are going to analyze physical processes in the framework of the 
molecular dynamics ensemble, which is the microcanonical ensemble plus the conservation of 
total momentum. Let us assume that at a given stage of the evolution of the system we can 
identify the biggest fragment according to $MST$ algorithm.

In order to study the time evolution of the thermodynamics of the biggest emiting source we 
define the local temperature $T_{loc}(t)$ as the velocity fluctuations around the mean radial 
expansion of the biggest source.  
So we first introduce the mean radial velocity of the biggest source, which is defined as

\begin{equation}
v_{rad}(t)=<\frac{1}{N_{BF}} \sum_{i=1}^{N_{BF}}\frac{ {\bf v}_{i}(t) \cdot {\bf r}_{i}(t)} 
{\left | {\bf r}_{i}(t) \right |} >_e
\label{eqflux}
\end{equation}

Where $N_{BF}$ indicates the number of particles of the biggest source at time $t$, and $< >_e$ 
denotes an average over all events. Both ${\bf v}_{i}(t)$ and ${\bf r}_{i}(t)$ are measured from 
the center of mass of the biggest source. Taking into account that collective motion should not 
be considered to calculate the temperature, we define the local temperature as $T_{loc}$:

\begin{equation}
T_{loc}(t) =\frac{2}{3 N_{BF}} \sum_{i=1}^{N_{BF}} \frac{m}{2}  
({\bf v}_{i}(t) - v_{rad}(t) \cdot {\bf \hat{r}}_{i}(t))^2 
\label{eqtemp}
\end{equation}

This definition is closely related to the one used in \cite{ale97} for the analysis of 
expanding systems. In that case we obtained the local temperature of the system by the 
following procedure: We divided our drops in concentric spherical regions, centered in the 
c.m of the system, of width $\delta r=2\sigma $.
The mean radial velocity of region $i$ in this case is:
\begin{equation}
v_{radshell}^{(i)}(t) = \frac{1}{N_i(t)} \sum_{ev} \sum_{j \in i} \frac{{\bf v}_{j}(t) \cdot {\bf r}_{j}(t)} {\left| {\bf r}_{j}(t) \right|}
\end{equation}
where the first sum runs over the different events for a given energy, the
second over the particles $j$ that belong, at time $t$, to region $i$;
${\bf v}_j$ and ${\bf r}_j$ are the velocity and position of particle $j$.
$N_i(t)$ is total number of particles belonging to region $i$ in all the
events. Then the local temperature, which will be called $T_{shell}^{(i)}$, is defined as:
\begin{equation}
T_{shell}^{(i)}=\frac 2 3 \frac 1{N_{i}} \sum_{j\in i}\frac 12m
     \left( {\bf v}_j-\frac{v_{rad}^{(i)}\cdot {\bf r}_j}{\left|
        {\bf r}_j\right| }\right) ^2
	\label{eqtemp2}
\end{equation}
where $N_i$ is the total number of particles in cell $i$ in all
events. 

The validity of Eq.\ref{eqtemp} and  Eq.\ref{eqtemp2} rely on the {\it conjecture} that the 
fragmenting system achieves local equilibrium (local equilibrium hypothesis). 
Whereas in the last case all particles in the 
inner shells should be equilibrated, the advantage of dealing with the local temperature 
(Eq.\ref{eqtemp}) relies on the fact that only 
the biggest source is assumed to have reached some degree of equilibration. In this paper we 
will focus mainly on this definition. If the biggest source is assumed to have reached 
equilibrium, the velocity distribution should follow:

\begin{equation}
f(v)=\rho (\frac{m \beta}{2\pi})^{3/2} e^{-\beta \frac{m}{2}(v-v_{rad})^2}
\label{eqmaxwell}
\end{equation}

A first approach on the study of the local equilibrium hypothesis (LEH) consists in analyzing the 
isotropy of the velocity fluctuations around the expansion. For this purpose we introduce 
the radial and transversal local temperatures:

\begin{equation}
T_{loc}^{rad} =\frac{2}{N_{BF}} \sum_{i=1}^{N_{BF}} \frac{m}{2}  
[({\bf v}_{i}(t) - v_{rad}(t) \cdot {\bf \hat{r}}_{i}(t)) \cdot {\bf \hat{r}}_{i}(t)]^2 
\label{tradial}
\end{equation}

\begin{equation}
T_{loc}^{tra} =\frac{2}{N_{BF}} \sum_{i=1}^{N_{BF}} \frac{m}{2}  
[({\bf v}_{i}(t) - v_{rad}(t) \cdot {\bf \hat{r}}_{i}(t)) \cdot {\bf \hat{r_{\perp}}}_{i}(t)]^2 
\label{ttransversal}
\end{equation}

A more profound analysis of the accuracy of the $LEH$ is the study of the velocity distribution 
function and its comparison with (Eq.\ref{eqmaxwell}). In order to perform this study we are to 
check the significance $S$ of the fit of the distribution numerically obtained with the 
Eq.\ref{eqmaxwell}. 
To address the question if both distributions are different we will perform perform a Pearson 
$\chi^2$ test \cite{numrec}. If the significance is high enough not to reject the hypothesis that
 both distributions differ we obtain as a by product the possibility of calculating the 
temperature as:

\begin{equation}
T_{max}=m \sigma^2
\end{equation}

Where $\sigma$ is the width of the velocity distribution. 

\section{The model}

The system under study is composed, as in previous works, by $N=147$ particles interacting 
via a truncated and shifted Lennard-Jones ($LJ$) potential, with a cut off radius $r_{cut}=3\sigma$. 
Energies are measured in units of the potential well ($\epsilon $), and the distance at which the 
$LJ$ potential changes sign ($\sigma$), respectively. The unit of time used is 
$t_{0}=\sqrt{\sigma^{2}m/48\epsilon }$.

The typical frequency of such potential is $\nu_0\sim 0.2 \frac{1}{t_0}$ . 
This defines a minimum time scale for the stability of an interacting system.
 
The equations of motion were integrated using the velocity Verlet algorithm, which preserves 
volume in phase space. We used an integration time step of $0.01 t_0$, and performed a  
microcanonical sampling every $1t_0$ up to a final time of $t=250t_0$.

This time scale was chosen because is has proven to be long enough as to the system to 
attain microscopic stability.

As stated in the previous section, we are mainly interested in the properties of fragments in 
configuration space. Then we are to analyze the $MD$ evolutions using the $MST$ algorithm.                                                                                                     
We define a fragmentation process when a source emits a {\emph stable} fragment of
at least $4$ particles, i.e. the $MST$ determination of fragments is complemented with a temporal
stability condition. Otherwise we will be facing an evaporation process. The evolutions are 
analyzed in an event-by-event basis and the times at which fragmentation takes place are determined.

\section{Numerical experiments}

Initial configurations are built as dense drops in $q-space$ (region C of the phase diagram 
\cite{phasediagram}). The total linear and angular momentum of the drops are removed and then, 
velocities are rescaled with a Maxwellian distribution so that the system has any desired 
value of total energy.  

We covered a broad range of energies, namely $-2.0\epsilon$,  $-1.0\epsilon$, 
$-0.5\epsilon$,  $-0.2\epsilon$, $0.0\epsilon$,  $0.2\epsilon$, $0.5\epsilon$,  
$1.0\epsilon$ and $2.0\epsilon$ (all reported energies are per particle). 
For each energy $1000$ events were calculated.

\subsection{Fragment mass distributions}

\begin{figure}[htbp]
   \setlength{\abovecaptionskip}{40pt}
   \centering
   \includegraphics[width=8cm,angle=0]{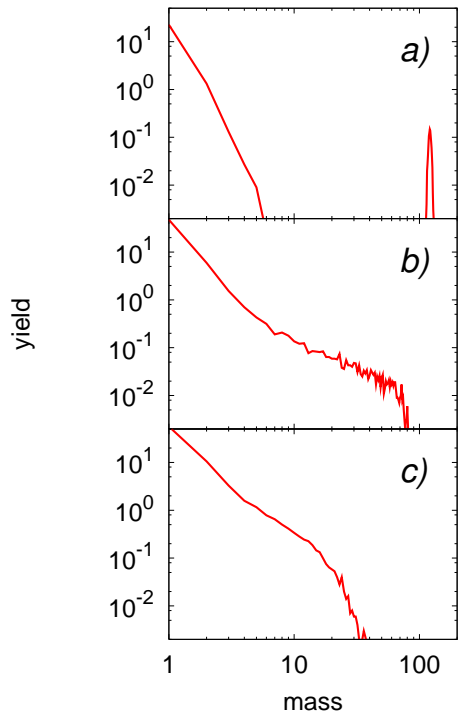}%
   \caption{ Asymptotic mass spectra. Energies $E=-2.0\epsilon, E=0.0\epsilon$, 
and $E=2.0\epsilon$ are displayed in panels $a)$, $b)$, and $c)$, respectively.}
   \label{spectra}
\end{figure}

The range of energies we used covers all regions of interest in multifragmentation. 
In Fig.~\ref{spectra}, it can be clearly seen that the asymptotic mass distributions go 
from "U-shaped", at low energies to exponentially decaying ones, at high energies. Between 
these shapes a power-law (panel $b)$ in Fig.~\ref{spectra} can be found. Technical details 
of the fitting procedure can be found elsewhere \cite{balencrit} but we would like to point out 
that the fitting procedure excludes the biggest fragment at each event, and also explicity 
excludes monomers, dimers and trimers.  

\begin{figure}[htbp]
   \setlength{\abovecaptionskip}{40pt}
   \centering
   \includegraphics[width=8cm]{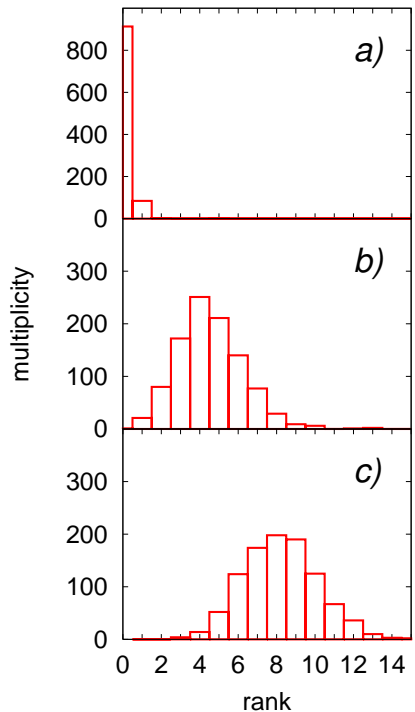}%
   \caption{Multiplicity distribution for $E=-2.0\epsilon, E=0.0\epsilon$, and $E=2.0\epsilon$ 
are shown in panels $a)$, $b)$, and $c)$, respectively.}
   \label{multiplicity}
\end{figure}

We have found interesting to study the multiplicity distribution of the number of times the 
system fragments for all events as a function of the energy 
of the system. In this case our approach is to follow the dynamics of the biggest source. At 
a given time $t_i=n t_0$ we identify the biggest $MST$ fragment and check for the 
particles that belonged to this fragment at $t_i$, and formed a smaller cluster at $t=(n+5)t_0$. 
We have chosen $5t_0$ as our threshold to accept the cluster as stable because that is of the 
order of the inverse of the natural frequency. In this way, we show in Fig.\ref{multiplicity} this 
distribution for energies per particle $E=-2.0\epsilon, E=0.0\epsilon$, and $E=2.0\epsilon$.
It can be easily seen that, in this range of energy, as the energy of the system is increased, 
the maximum of the multiplicity distribution shifts towards higher values while the width of 
the distribution increases.

\subsection{Characteristic times}

\begin{figure}[htbp]
   \setlength{\abovecaptionskip}{40pt}
   \centering
   \includegraphics[width=8cm]{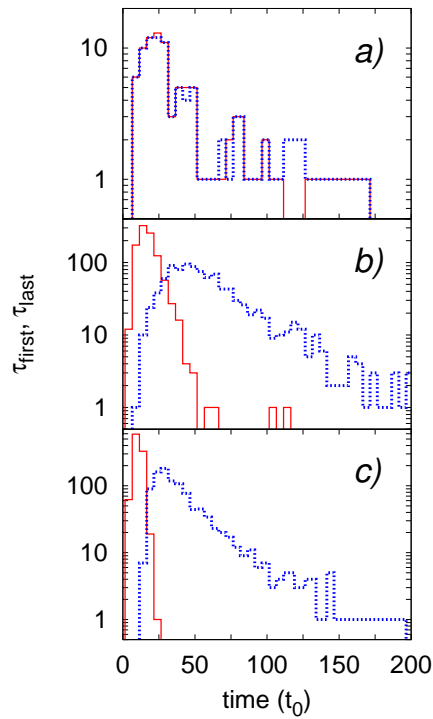}%
   \caption{(Color online) Distribution of first (full line) and last (dotted line) time of emission. 
Energies: $E=-2.0\epsilon, E=0.0\epsilon$, and $E=2.0\epsilon$ are displayed in panels 
$a)$, $b)$, and $c)$, respectively.}
   \label{firstandlast}
\end{figure}

Some discussion regarding the nature of the fragmentation process has focused on the distinction
 between simultaneous and sequential kind of phenomena. In our calculation the analysis of these
 kind of things are straightforward. Anyway, there are different characteristic times which can 
be explored. First we will look at the first and last time of emission. This means that we 
record for each of our evolutions the time at which the first fragmentation takes place according
 to the aboved-mentioned definition of an emission process. The result of that calculation are 
displayed in Fig.\ref{firstandlast}. 

\begin{figure}[htbp]
   \setlength{\abovecaptionskip}{40pt}
   \centering
   \includegraphics[width=8cm]{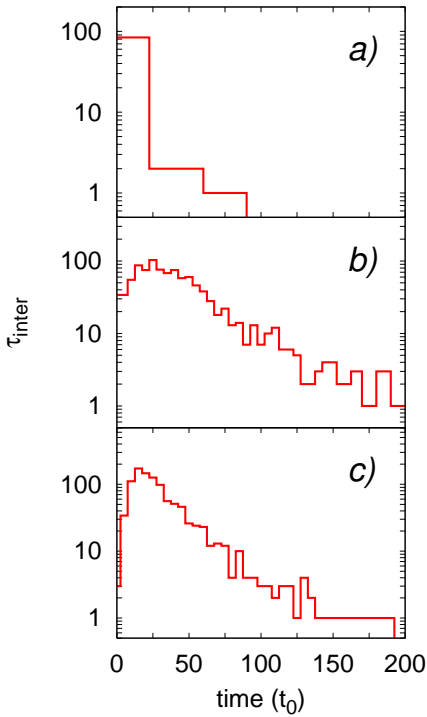}%
   \caption{Distribution of lapse of time between the first and last emissions. 
Energies: $E=-2.0\epsilon, E=0.0\epsilon$, and $E=2.0\epsilon$ are displayed in 
panels $a)$, $b)$, and $c)$, respectively.}
   \label{firstandlast2}
\end{figure}

\begin{figure}[htbp]
   \setlength{\abovecaptionskip}{40pt}
   \centering
   \includegraphics[width=8cm]{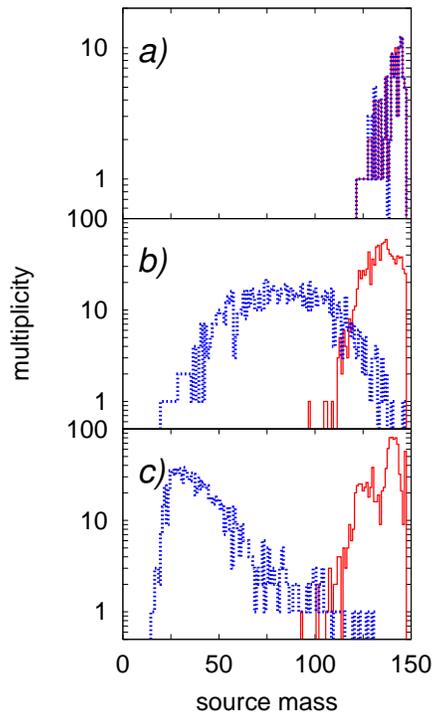}%
   \caption{(Color online) Distribution of mass of the emitting sources for the first (full line) 
and last (dotted line) emissions. Energies: $E=-2.0\epsilon, E=0.0\epsilon$, and $E=2.0\epsilon$ are 
displayed in panels $a)$, $b)$, and $c)$, respectively.}
   \label{sourcedist}
\end{figure}

We can see that at high energies ($E=2.0\epsilon$) the
 distributions of times are narrower that at lower energies and does not seem to fit the view
 of a simultaneous process in configuration space (This issue will be further analyzed in 
Section \ref{redu}). Another view of the same data is shown in Fig.\ref{firstandlast2} in which 
what we show is the corresponding lapse of time between the first and last emission for each 
event. Once again, the distribution gets broader as the energy gets lower (the comparison should
 be restricted to events corresponding to $E=0.0\epsilon$ and $E=2.0\epsilon$ because of the 
very few emission processes that are found at $E=-2.0\epsilon$. In order to further illustrate
 the kind of process we are facing we show in Fig.\ref{sourcedist} the distribution of the mass 
of the emitting sources for the times of first and last emissions. 

The most relevant issue of this figure is the presence, in panel $b)$, of a broad range of 
masses at which the last emission takes place, since the value of energy ($E=0.0\epsilon$) 
correspond to a power-law mass distribution.
 
The reason why we stated above that these results do not seem to fit the simultaneous emission
 picture without assuring it, is because we can also explore the possibility of occurrence of
 massive emission events. Even though emission events can be found in a rather large scale of 
time one may wonder if massive emission events have a definite characteristic time which would 
sustain the simultaneous picture.

\begin{figure}[htbp]
   \setlength{\abovecaptionskip}{40pt}
   \centering
   \includegraphics[width=8cm,angle=0]{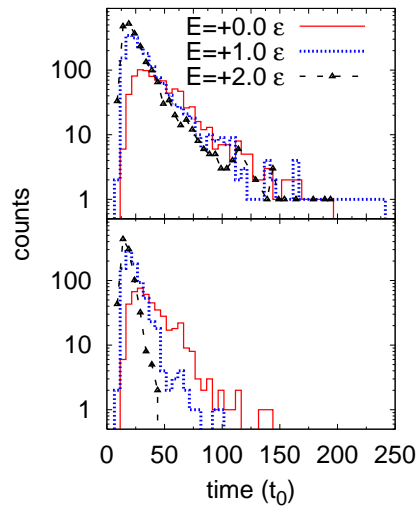}%
   \caption{(Color online) Distribution of emission times of massive fragments using 
criterium $1$ (top) and criterium $2$ (bottom); see text for details.}
   \label{massive}
\end{figure}

We have used two definitions of "massive emission events (MEE)". The first one considers a MEE 
if the mass of the emitted fragment is at least $30\%$ of the total mass of the biggest source 
at the time of emission. The second one states that a MEE has taken place if the mass of the 
biggest emitted fragment is at least composed of $40$ particles. The results of such 
calculations are shown in Fig.\ref{massive}.       

It is immediate that massive emission, according to definition $1$, can take place at any time 
during the evolution, while when using the second definition (which is more restrictive) there 
are barely no massive events for times greater than $t \sim 100 t_0$. Also notice that the 
lower value of energy we used ($E=-2.0\epsilon$) is not present because no massive 
fragmentations were found.

In conclusion, when gathering all these results together we reach the conclusion that for the 
energies displayed in these figures the process can be viewed as a mixture of sequential and
 simultaneous break up.

\subsection{Temperatures of the emitting sources}

\begin{figure}[htbp]
   \setlength{\abovecaptionskip}{10pt}
   \centering
   \includegraphics[height=8cm,angle=270]{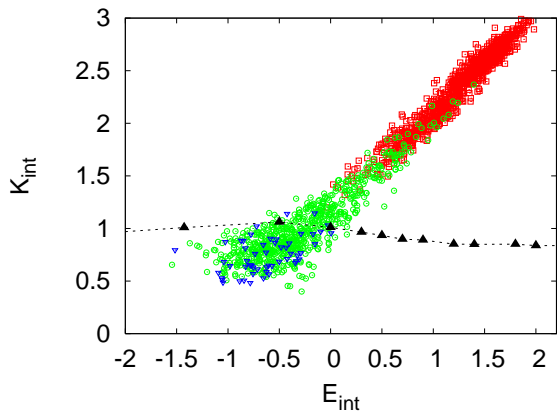}%
   \caption{(Color online) Internal kinetic energy of the emitting source as a function of its total 
internal energy for three ranks. Empty (red) squares denote $1st$ emission process. Empty (green) 
circles denote the $8th$ emission process, and empty (blue) triangles indicate the $12th$ emission 
process. Full (black) triangles stand for the caloric curve of the expanding system. }
   \label{kintvse}
\end{figure}

It is natural to think that as the sources emit fragments they will undergo a cooling process. 
We should keep in mind that we are trying to analyze things in an event-by-event basis, without 
performing averages that would obscure the picture. In order to gain knowledge of the 
microscopic view of the fragmentation phenomena, we have calculated the evolution of the 
temperature, for a given energy and for each time step in which a fragmentation event 
takes place. The temperature is the one of the emitting source in the time
 step prior to the one in which the emission takes place. We then plot all this information in 
a single graph (see Fig.\ref{kintvse}).  

In Fig.\ref{kintvse} we show the result for the analysis of 1000 events at an initial energy of
 $E=2.0 \epsilon$. In order to make the figure more readable we are only showing the 
above-described temperature for the first, last and maximum multiplicity fragmentation events. 
It is clearly seen that the internal kinetic energy is linear with the total internal energy and 
displays a cooling behavior. This overall behavior will not change if we consider not only 
fragmentation events but also evaporation events. It is interesting to notice that if one is to 
calculate temperatures from the analysis of the emitted light fragments (monomers, dimers, trimers)
 one would be sampling a source that starting from a rather high temperature, cools down 
monotonically. So the corresponding temperature should display a maximum at early times and 
 decrease later on \cite{natowitz}. This is not a problem at all because the system 
is out of equilibrium and as such the kind of measurements we are referring to at this point 
do not correspond to stable systems. What might be improper is to talk about temperature 
instead of "effective temperature". A more quantitative calculation is under progress and will 
be comunnicated shortly.

\subsection{The role of radial flux}

\begin{figure}[htbp]
   \setlength{\abovecaptionskip}{10pt}
   \centering
   \includegraphics[height=8cm,angle=270]{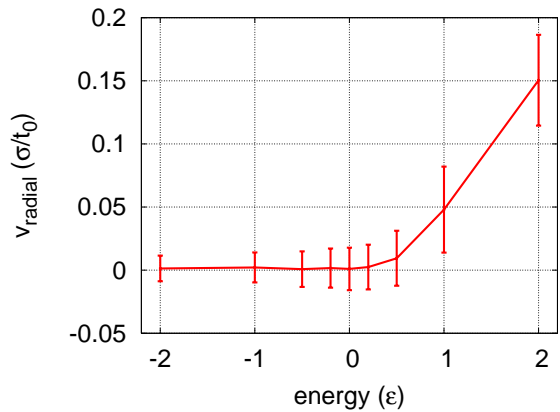}%
   \caption{ Radial flux as a function of energy at time of fragment formation $\tau_{ff}$}
   \label{flux}
\end{figure}

In a series of previous works \cite{noneqfrag} we have shown that, if the presence of collective 
(nonthermal) motion is not taken into account one find a wrong result: The presence of a vapor-branch 
in the caloric curve. Only when the radial flux is properly incorporated in the definition of the 
temperature one find that the nonequilibrium process of multifragmentation 
appears as an almost constant temperature region at high energies. In particular, in 
paper \cite{ale97b} we have calculated the time of fragment formation of this system, 
(using the ECRA phase-space method). The values of the mean radial velocity calculated for 
the biggest source in each event according to Eq.\ref{eqflux} are shown in Fig.\ref{flux}.  

\begin{figure}[htbp]
   \setlength{\abovecaptionskip}{10pt}
   \centering
   \includegraphics[height=8cm,angle=270]{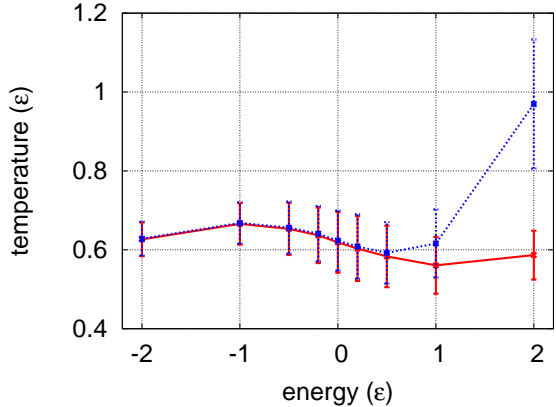}%
   \caption{(Local temperature) Local temperature of the biggest source at $\tau_{ff}$ 
(straight red line) and "fake temperature" (dotted blue line). The fake temperature is calculated 
from the total kinetic energy of the biggest source without taking into account the importance 
of the radial flux}
   \label{tlocalandtint}
\end{figure}

In Fig.\ref{tlocalandtint} we show the local temperature of the biggest source at time of fragment
 formation, as a function of the total energy (full line) The dotted line is included in order to
 emphazise the effect of not removing the radial collective motion when calculating caloric curves.
 As explained in Section \ref{sectemp} temperatures were calculated from velocity fluctuations 
around the collective motion.

\begin{figure}
\centering
\includegraphics[height=8cm,angle=270,clip=]{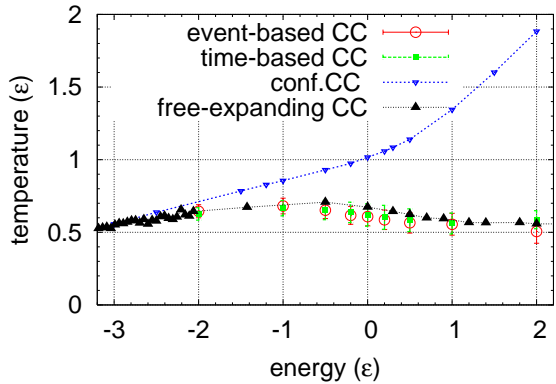}%
\caption{(Color online) In this figure we show a comparison of caloric curves obtained by 
different thermometers. Empty circles correspond to an event-based temperature. Filled squares 
correspond to the temperature of the biggest source at time of frament formation. Triangles up
denote the local temperature ($T_{shells}$) and the dotted line with down triangles corresponds to 
the caloric curve obtained for a constrained system with a density $\rho=0.17\sigma^{-3}$.}
   \label{caloriccurves}
\end{figure}

In order to see how the picture emerging from a phase space analysis relates to the main topic of 
this paper i.e. the analysis in configuration space, we proceeded to calculate the temperature of
 the emitting sources at the stage of the evolution which correspond to the maximum of the 
multiplicity distribution (for example, looking at Fig.\ref{multiplicity}b) the maximum of 
the distribution turns out to be rank=$4)$. From this kind of analysis we get the empty circles 
of Fig.\ref{caloriccurves}. It can be easily seen that both approaches give the same temperature.

\subsection{Local Equilibrium Hypothesis $(LEH)$}

\begin{figure}[htbp]
   \setlength{\abovecaptionskip}{40pt}
   \centering
   \includegraphics[width=8cm]{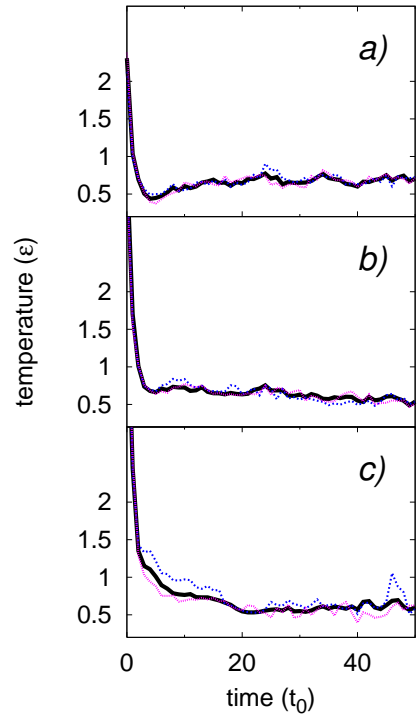}%
   \caption{(Color online) Radial (dotted line), transversal (thin line), and local 
(full line) temperatures of the biggest source for $E=-2.0\epsilon$ (panel $a)$), 
$E=0.0\epsilon$ (panel $b)$), and $E=2.0\epsilon$ (panel $c)$)}
   \label{tlocrad}
\end{figure}

In previous sections we have made use of the hypothesis of local equilibrium. 

In order to check this hypothesis we have performed the following calculation. Because in a local
 equilibrium scenario the fluctuations of velocity should be the same for the radial and 
transversal directions, we evaluated Eqs.~\ref{tradial} and ~\ref{ttransversal}, 
for a single event at the three energies chosen as examples in this work. In each of the three 
panels of Fig.\ref{tlocrad}, each corresponding to a different total energy (see caption for 
details), we show the three definitions of temperature above-mentioned. It is immediate that all 
of them are essentially the same, except for very early stages of the evolution (times much 
shorter than the time of fragment formation). This suggest that the the $LEH$ at time of fragment
 formation seems plausible for all energies. 

\subsection{Maxwellian distribution of velocities}

\begin{figure}
\centering
\includegraphics[width=8cm,clip=]{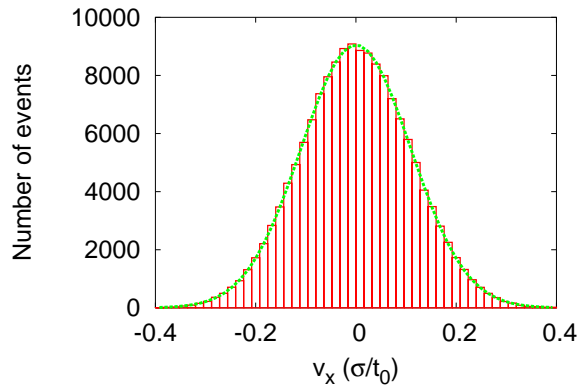}%
\caption{(Color online) Velocity distribution (histogram) and Maxwellian fit (dotted line) 
for $E=+2.0\epsilon$}
\label{maxwellian}
\end{figure}

In this section we will show that the velocity distribution of the biggest source at time of 
fragment formation is indeed Maxwellian. Moreover, the temperature obtained from the standard 
deviation of the velocity distribution is almost exactly the same of Eq.\ref{eqtemp}, showing 
the consistency of our numerical studies. In Fig.\ref{maxwellian} we show the histogram of the 
velocity distribution at time of fragment formation and its Maxwellian fit for $E=2.0 \epsilon$. 

In adittion to the excellent agreement between the velocity distributions and their fits 
we performed a Pearson $\chi^2$ test, trying to reject the hypothesis that the distribution 
of velocities and that corresponding to the Maxwellian fit differ. We found a significance ($S$) 
above $0.25$ for all cases, showing that, even with a very low confidence level like $CL=0.80$ 
we could reject the hypothesis that both distributions differ (i.e. the velocity distribution is 
indeed Maxwellian).
                                                                                                     \begin{table}[htbp]
\begin{center}
\centering
\caption{Comparison of thermometers and statistical significance of the Maxwellian fit}.
\begin{ruledtabular}
\begin{tabular}{|l|l|l|l|l|l|}
 $E (\epsilon)$  &  $T_{local} (\epsilon)$       &  $T_{fit} (\epsilon)$ &  $T_{event} (\epsilon)$ 
& $T_{shell} (\epsilon)$ & {\it S}\\ \hline
 $-2.0$  &  $0.627\pm 0.043$  &  $0.623$   &  $0.645$     &  $0.647$    & $0.86$       \\ 
 $\quad 0.0$  &  $0.619\pm 0.077$  &  $0.630$   &  $0.608$     &  $0.674$    & $0.82$       \\ 
 $\quad 2.0$  &  $0.587\pm 0.062$  &  $0.575 $  &  $0.506$     &  $0.558$    & $0.78$       \\ 
\end{tabular}
\end{ruledtabular}
\end{center}
\label{tabletemp}
\end{table}

In Table.$I$ we show a cross comparison of different thermometers. In addition to the already 
presented results we add the temperature obtained from the Maxwellian fit of the velocity 
distribution, and we also show the obtained statistical significance.

\subsection{Reducibility}
\label{redu}

Not long ago, it has been proposed by Moretto and co-workers \cite{moretto1,moretto2} that the 
complex process of fragment emission could be described in terms of a binomial distribution. This
 approach rests on the assumption that a single transition probability $p$ is capable of describing
 the emission process when no regard is pay to the mass or composition of the emitted fragments.
 In this way, the probability of emitting $n$ fragments in a series of $m$ trials should 
follow the well-known binomial distribution:

\begin{equation}
P^m_n = \frac{m!}{n!(m-n)!} p^n (1-p)^{m-n}
\label{eqbinomial}
\end{equation}

\begin{figure}[htbp]
   \setlength{\abovecaptionskip}{40pt}
   \centering
   \includegraphics[width=10cm]{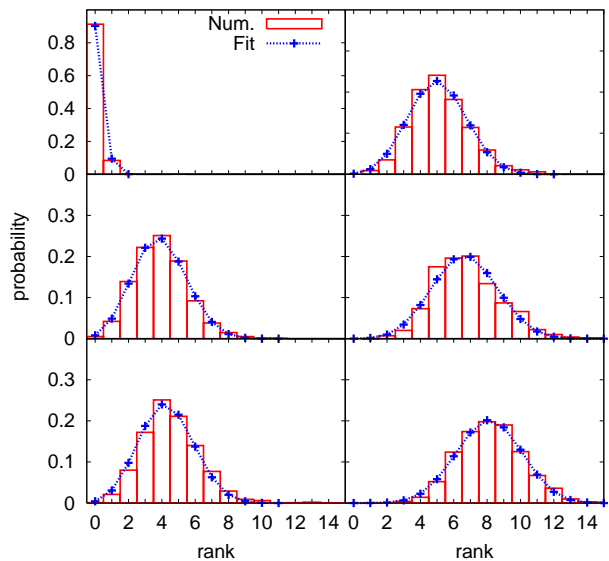}%
   \caption{(Color online) Multiplicity distribution (histograms) and binomial fit (dotted line). 
For energies (from top to bottom and left to right): $E=-2.0\epsilon, E=-0.2\epsilon$, 
$E=0.0\epsilon, E=0.2\epsilon$, $E=1.0\epsilon$, and $E=2.0\epsilon$}
   \label{binomial}
\end{figure}

In which $m$ stands for the number of "trials", while $n$ stands for the number of successes.
 Following \cite{moretto1}, we associate the parameter $m$ to the maximum multiplicity for each 
energy and $p$ stands for the transition probability. In Fig.\ref{binomial} we show the 
result of such an analysis. The quality of the resulting fit is remarkable indeed, specially when one 
consider that we are facing an out of equilibrium non-simultaneous process while the very nature
 of the binomial process requires a constant value of $p$ for the whole emission process.

\begin{figure}[htbp]
   \setlength{\abovecaptionskip}{0pt}
   \centering
   \includegraphics[height=8cm,angle=270]{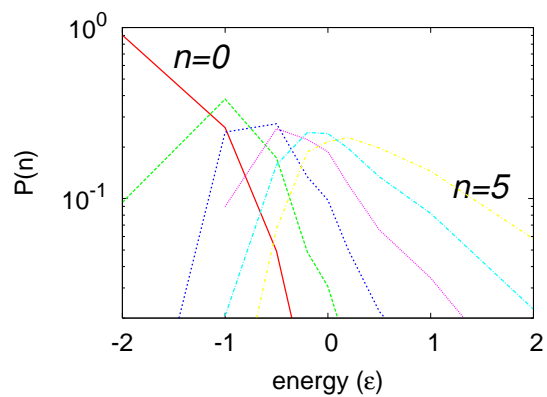}%
   \caption{ (Color online) Probability of emiting 0 (red), 1 (green), 2 (blue), 3 (violet), 
4 (cyan), 5 (yellow) fragments as a function of energy}
   \label{multiproba}
\end{figure}

In order to further illustrate the accuracy of this approach we show in Fig.\ref{multiproba} 
the calculated probability of emitting $n$ fragments during the entire process as a function
 of the total energy, whith $n=0-5$. $P(n)$ is calculated asssuming a binomial distribution 
( Eq.~\ref{eqbinomial}) with the value of $p$ obtained from the best fit. 

We will not further analyze the implications of this binomial fit because we have not been able 
to calculate microscopically the associated transitions barriers yet.

\section{Conclusions}

In this work we have analyzed the dynamics and thermodynamics of the emission process in a 
classical fragmenting system. We have focused on the evolution of the biggest source, and the 
corresponding emitted fragments. We have shown that the emission process can not be cast into a 
sequential nor into a simultaneous scenario, because even though there are many events in which 
fragments are emitted sequentially the phenomenon of massive emission is frequent enough as to 
forbid the characterization as a purely sequential one. We have also shown that when looking at 
fragments well defined in configuration space, the process of emission can not be cast into a 
isothermal one, i.e. temperatures are time-dependent. Moreover, there is a 
time-dependence of the size of the emitting source. As such, 
standard thermodynamical models which fix temperature and volume do not seem to be appropiate to
 reveal the true nature of the phenomenon under analysis. We have also shown, by analyzing the 
temperature of the source at the stage of most probable multiplicity, that it is possible to 
recover the caloric curve already obtained in the frame of phase-space analysis. If we recall 
Table $I$ we see
 that temperature can be determined according to a few reasonable definitions, which provides in 
all cases consistent results.

Two other interesting results have been obtained as a by-product of these calculations: First 
we should like to mention that the velocity distribution functions of the particles that form 
the biggest source are remarkably Maxwellian for times larger than the time of fragment formation
 after the collective expanding mode is removed. Second we have found that the probability of 
fragment emission (summed-up over all sizes) is amazingly binomial, suggesting a constancy of 
the transition amplitudes. We do not have a microscopic description of this constancy right now 
and we are currently working on it.

We hope that these findings will encourage the development of new, more accurate and realistic 
models to describe nuclear multifragmentation.

\section*{Acknowledgements}

We acknowledge partial financial support from the University of Buenos Aires via grant $X308$. 
We also acknowledge partial financial support from CONICET (via grant $PIP 2304$). C.O.D. is a 
member of the Carrera del Investigador (CONICET). M.J.I. is a fellow of UBA.


\bibliographystyle{ieeetr}
\bibliography{emission}

\end{document}